%% file: soft-qqQQ.tex
\documentclass[12pt]{article}
\usepackage{jheppub}
\pdfoutput=1
\usepackage{amsmath,amssymb}
\usepackage{mathrsfs}

\usepackage{graphicx}
\usepackage{slashed}
\usepackage{bbold}
\usepackage{braket}
\usepackage{float}
\usepackage[most]{tcolorbox}
\usepackage{pdfpages}
\usepackage{hyperref}
\usepackage{float}
\usepackage{url}

\usepackage{tikz}
\usetikzlibrary{arrows.meta, decorations.pathmorphing, shapes, positioning,calc}
\usetikzlibrary{decorations.markings}
\usepackage{circuitikz}

\tikzstyle directed=[postaction={decorate,decoration={markings, mark=at position .65 with {\arrow[arrowstyle]{stealth}}}}]

\definecolor{GWmerger}{RGB}{112,176,66}
\definecolor{GWringdown}{RGB}{237,122,38}
\definecolor{internationalorange}{rgb}{1.0, 0.31, 0.0}
\newcommand\cdotnew{\!\cdot\!}
\DeclareMathOperator{\tr}{tr}  

\begin{document}

\title{Tree-level soft emission for two pairs of quarks}
\author[Q]{Xinguang Chen}
\author[C,D]{and Zhengwen Liu}

\affiliation[Q]{School of the Gifted Young, University of Science and Technology of China,\\
Hefei 230026, China}
\affiliation[C]{School of Physics \& Shing-Tung Yau Center, Southeast University, \\
Nanjing 210096, China}
\affiliation[D]{Niels Bohr International Academy, Niels Bohr Institute, University of Copenhagen,\\
Blegdamsvej 17, 2100 Copenhagen \O, Denmark}

\emailAdd{chenxinguang@mail.ustc.edu.cn, zhengwen.liu@seu.edu.cn}

\abstract{We compute the tree-level current for the emission of two soft quark-antiquark pairs in a hard scattering. We also compute the square of this current and discuss the resulting color correlations, featuring dipole correlations and three-parton correlations. This object is essential for analyzing the infrared singularities at next-to-next-to-next-to-next-to-leading-order (N$^4$LO) predictions in perturbative QCD.}

\arxivnumber{2411.08795}

\setcounter{tocdepth}{1}
\maketitle

\section{Introduction}

In gauge theory such as quantum chromodynamics (QCD), on-shell scattering amplitudes exhibit a universal factorization behavior in specific kinematic regimes, such as the soft and collinear limits. Notably, when one or more massless partons become {\it soft}, the full scattering amplitude factorizes into a product of a process-independent soft factor, represented by an operator acting on the color space, and a lower-point scattering amplitude with the soft particles removed \cite{Low:1958sn,Yennie:1961ad,Weinberg:1965nx,Weinberg:1964ew,Burnett:1967km,Jackiw:1968zza,Bassetto:1983mvz,Berends:1988zn}. Explicit analytic expressions for this universal factorization are vital for constructing so-called subtraction schemes, which play a crucial role in managing and canceling infrared singularities in cross sections, see e.g.\,\cite{Catani:1996vz,Frixione:1995ms}. Moreover, the soft factorization of scattering amplitudes, often referred to as soft theorems in literature, sometimes can be useful to reveal new mathematical structures of S-matrices, shedding light on the hidden symmetries underlying quantum field theory. Therefore, a refined understanding of the soft factorization of scattering amplitudes is of fundamental importance for performing precision computations of physical observables as well as for exploring novel structures in quantum field theory.

\vskip 5pt
Motivated by its significance in both phenomenological and theoretical studies, the soft factorization of gauge theory amplitudes has been extensively investigated. For the emission of a single soft gluon, the explicit factorization formulas for both unsquared and squared amplitudes have been established at tree \cite{Low:1958sn,Yennie:1961ad,Weinberg:1965nx,BASSETTO1983201,Proceedings:1992fla}, one-loop \cite{Bern:1998sc,Bern:1999ry,Catani:2000pi} and two-loop \cite{Li:2013lsa,Duhr:2013msa,Dixon:2019lnw} levels in the strong coupling constant $\alpha_s$ in perturbative QCD. For the emission of double soft partons (tow gluons or a quark-antiquark pair), the soft currents have been known at tree \cite{Catani:1999ss,Campbell:1997hg,Czakon:2011ve} and one-loop \cite{Zhu:2020ftr,Catani:2021kcy,Czakon:2022dwk} levels. Recently, the tree-level current for the emission of triple soft partons, including three gluons ($ggg$) and a soft quark-antiquark pair plus a soft gluon ($q\bar{q}g$), has also be worked out \cite{Catani:2019nqv,DelDuca:2022noh,Catani:2022hkb}, see also \cite{Haindl:2022hcb} for a recent review. Among these analytic results, the tree-level double soft parton currents and the tree-level and one-loop single gluon soft currents have been essential ingredients in building subtraction schemes at next-to-leading order (NLO) and next-to-next-to-leading order (NNLO). The other higher-order results, in turn, are pivotal for next-to-next-to-next-to-leading order (N${}^3$LO) computations in QCD.

\vskip 5pt
Beyond N${}^3$LO, the single soft gluon current has recently been computed to three-loop level in QCD and maximally supersymmetric Yang-Mills theory \cite{Herzog:2023sgb,Chen:2023hmk}. The present work aims to compute the emission of four soft partons at the tree level. We concentrate on in this work the current for the soft emission of two quark-antiquark pairs. They will be essential for understanding the infrared singularities at next-to-next-to-next-to-next-to-leading-order (N${}^4$LO) QCD computations.

\vskip 5pt
The remainder of this paper is structured as follows. In section~\ref{sec-eikonal-approximation}, we briefly review the eikonal approximation formalism that is our main computational method, along with a concise introduction to the color space representation of scattering amplitudes to clarify concepts and streamline notation. We then present the primary results of this paper: the tree-level current for the emission of two soft quark-antiquark pairs in section \ref{sec-soft-qqqq-current}. We construct the current using the eikonal formalism, and then compute its square and analyze resulting color correlations. We conclude with a summary in section~\ref{sec-conclusion}.

\section{The eikonal approximation}
\label{sec-eikonal-approximation}

In this section, we present a brief review of the eikonal approximation formalism of the tree-level current for the emission of soft partons.

\subsection*{The color space formalism}
In order to keep notations compact, we find it useful to employ the colour-space formalism \cite{Catani:1996jh,Catani:1996vz}, in which scattering amplitudes can be represented as a vector that can be expanded into an orthonormal basis in color $\otimes$ spin space
\begin{align}\label{def-color-spin-space}
    \Ket{\mathcal{M}_{f_1\ldots f_n}(p_1,\ldots,p_n)} = \Ket{c_1\ldots c_n} \otimes \Ket{\lambda_1\ldots \lambda_n} \mathcal{M}^{c_1\ldots c_n;\lambda_{1}\ldots \lambda_n}_{f_1\ldots f_n}(p_1,\ldots,p_n)\,,
\end{align}
where $|c_1\ldots c_n\rangle$ and $|\lambda_1\ldots \lambda_n\rangle$ are an orthonormal basis for color and spin respectively, and the coefficient
\begin{align}
    \mathcal{M}^{c_1\ldots c_n;\lambda_{1}\ldots \lambda_{n}}_{f_1\ldots f_n}(p_1,\ldots,p_n)
    := \big(\Bra{c_1\ldots c_n} \otimes \Bra{\lambda_1\ldots \lambda_n}\big) \Ket{\mathcal{M}_{f_1\ldots f_n}(p_1,\ldots,p_n)}
 \end{align}
is the amplitude of $n$ particles carrying momenta $\{p_1,\ldots,p_n\}$, flavors $\{f_1,\ldots,f_n\}$, colors $\{c_1,\ldots,c_n\}$ and spin indices $\{\lambda_1,\ldots,\lambda_n\}$.
In this work we define the momenta as well as spins (helicities) by assuming the external particles (at least partons) are outgoing for convenience.
Below, unless necessary, we will omit the dependence on momenta for simplicity.
The squared matrix element (summed over spin and colour indices of the external particles) can be written as
\begin{align}\label{amp-vec-sq}
\left|\mathcal{M}_{f_1\ldots f_n}\right|^2 
\equiv \langle\mathcal{M}_{f_1\ldots f_n} | \mathcal{M}_{f_1\ldots f_n} \rangle  
&=
\sum_{\substack{c_1,\ldots,c_n \\ \lambda_1,\ldots,\lambda_n}}\big(\mathcal{M}_{f_1\ldots f_n}^{c_1\ldots c_n; \lambda_1\ldots \lambda_n}\big)^\dagger
\mathcal{M}_{f_1\ldots f_n}^{c_1\ldots c_n; \lambda_1\ldots \lambda_n}
\,.
\end{align}

In this formalism, the color charge carried by a parton can be represented as an abstract operator that acts on the color space.
These color charge operators obey the $\mathrm{su}(N_c)$ color algebra
\begin{align}\label{color-Lie-algebra}
    [\mathbf{T}_j^a, \mathbf{T}_l^b] = if^{abc}\mathbf{T}_j^c\,\delta_{jl},
\end{align}
where $f^{abc}$ with $(1 \leq a,b,c \leq N_c^2{-}1)$ is the structure constant of the color algebra.
For the parton $i$ one has
\begin{align}
    {\bf T}_i^a |c_1 \cdots c_n\rangle = |c_1 \cdots c'_i \cdots c_n\rangle\,{\bf T}_{c'_ic_i}^a\,,
\end{align}
where ${\bf T}^a_{c'_i c_i} = i f^{c'_iac_i}$ (the adjoint representation of the color algebra) if the parton $i$ is a gluon, ${\bf T}^a_{c'_i c_i} = t^a_{c'_ic_i}$ or ${\bf T}^a_{c'_i c_i} = - t^a_{c_ic'_i}$ ($1 \leq c_i, c_i' \leq N_c$, fundamental or anti-fundamental representation) is a quark or anti-quark.
In this way, partons in various different representations of the color gauge group can be represented in a unified way.
It is direct to verify that $C_i  = \mathbf{T}_i^2 \equiv \mathbf{T}_i^a \mathbf{T}_i^a$ is a Casimir operator: $C_i = C_A$ if the parton $i$ is a gluon and $C_i = C_F$ if the parton $i$ is a quark, where
\begin{align}\label{Casimir-operator}
C_A = N_c, \quad
C_F = T_F {N_c^2-1 \over N_c},\quad\text{with}~~
\tr(t^a t^b) = T_F\delta^{ab} = {1\over 2} \delta^{ab}.
\end{align}

By definition \eqref{def-color-spin-space}, any amplitude vector $\Ket{\mathcal{M}}$ is a color-singlet, the color conservation can be expressed as
\begin{align}\label{color-conservation}
\sum\nolimits_{i} \mathbf{T}_i^a \Ket{\mathcal{M}} = 0
\quad\text{or simply}\quad
\sum\nolimits_{i} \mathbf{T}_i^a = 0.
\end{align}

\subsection*{The soft current}

An on-shell scattering amplitude may exhibit a universal factorization form in the limit where one or more massless external particles become \textit{soft}, i.e., their energies approach zero.
To be more precise, considering the scattering of $n$ {\it hard} particles with momenta $p_i$ and $m$ \textit{soft} particles with momenta $q_i$, the leading singular behavior of the scattering amplitude at the tree level is captured by the following factorization formula:
\begin{align}\label{soft-factorization-general}
    \mathscr{S}_{q_1\cdots q_m}&\Ket{\mathcal{M}_{n+m}(p_1,\ldots,p_n; q_1,\ldots,q_m)}
    \nonumber\\[0.5 em]
    &\quad = (\mu^{\epsilon}g_s)^m\, \mathbf{J}(q_1,\ldots,q_m; p_1,\ldots,p_n) \Ket{\mathcal{M}_n (p_1,\ldots,p_n)},
\end{align}
where the symbol $\mathscr{S}_{q_1\cdots q_m}$ signifies retaining only the leading singular terms in the soft limit, $q_i\to 0$. In formula \eqref{soft-factorization-general}, $g_s$ is the strong coupling constant, $\mu$ is the scale introduced by dimensional regularization, and the analysis is performed within the Conventional Dimensional Regularization (CDR) scheme in $D = 4 - 2\epsilon$ dimensions in this work. The scattering amplitude $\Ket{\mathcal{M}_n}$ on the right-hand side is obtained by simply excluding the $m$ soft particles from the full amplitude $\Ket{\mathcal{M}_{n+m}}$ on the left-hand side. The function $\mathbf{J}$, dependent on both soft and hard particles, characterizes the leading divergent behavior of the amplitude in the soft limit.

\vskip 5pt
The eikonal approximation formalism offers a systematic method for constructing soft currents.
The most intuitive and constructive idea is to study the emission of a single soft gluon at tree level.
A key observation is that the leading contribution arises only from diagrams where the soft gluon is emitted from an external hard parton line,
\begin{align}\label{eikonal-vertex}
\begin{aligned}
\begin{tikzpicture}[scale=1]
  \fill[black,opacity=0.2] (0,0) circle (8pt);
  \draw[line width=0.8pt] (0,0) circle (8pt);
  \draw[decorate, decoration={coil, amplitude=1.9pt, segment length=2.9pt,aspect=0.9pt}, line width=0.6pt, color=internationalorange] (20pt,0) -- (34.1pt,17.8pt); 
  \draw[line width=0.9pt] (0:8pt) -- (0:43pt);  
\end{tikzpicture}
\end{aligned}
= 
    \mathbf{J}_i^{a;\mu}(p_i,q) = \mathcal{S}_i^\mu(q)\mathbf{T}_i^{a}
    \quad\text{with}\quad
    \mathcal{S}_i^\mu(p_i,q) \equiv \frac{p_i^\mu}{p_i\cdot q}\,.
\end{align}
This is well-known eikonal vertex. Here, $q$ and $a$ denote the momentum and color index of the soft gluon, respectively, while $i$ labels the hard parton line with momentum $p_i$ and $\mathbf{T}_i^a$ is the corresponding color charge operator. By QCD theory itself, quarks cannot be emitted directly from a hard line. However, a quark-antiquark pair can be produced from an intermediate soft gluon, which may be connected to a hard line via the eikonal vertex or other internal lines. Using this, along with other Feynman rules, one can systematically construct the leading current for the emission of any number of soft partons.

\vskip 5pt
The most representative example is the tree-level current for the emission of a single soft gluon.
Using the eikonal vertex \eqref{eikonal-vertex}, summing over all external hard (QCD) lines immediately gives the full soft current \cite{BASSETTO1983201,Proceedings:1992fla}
\begin{align}
    \mathbf{J}^{a;\mu}_g(q) = 
    \sum_{i=1}^n \mathbf{J}^{a;\mu}_i(q) =
    \sum_{i=1}^n
    \mathcal{S}_i^\mu(q)\mathbf{T}_i^{a}\,,
    \label{eq:single_soft}
\end{align}
or in the vector form in color-spin space
\begin{align}\label{J-single-gluon-vec}
    {\bf J}_g(q) = |a\rangle\otimes|\lambda\rangle\,{\bf J}^{a;\lambda}_g(q)
    = |a\rangle\otimes|\lambda\rangle\,\epsilon^\lambda_\mu(q,r)\,{\bf J}^{a;\mu}_g(q)\,,
\end{align}
where $r$ is a lightlike reference vector with $q\cdot r\neq0$.
Here we suppress the hard momenta in the current function $\mathbf{J}$ for brevity. As mentioned, a pair of massless quark and antiquark  can be emitted from an internal soft gluon line that originates from an external hard parton
\begin{align}\label{eikonal-vertex-qq}
\begin{aligned}
\begin{tikzpicture}[scale=1]
  \fill[black,opacity=0.2]  (0,0) circle (8pt);
  \draw[line width=0.8pt] (0,0) circle (8pt);
  \coordinate (d) at  (20pt,16pt);
  \coordinate (e) at  (28pt,7pt);
  \coordinate (f)  at  (43pt,14pt);
  \draw[decorate, decoration={coil, amplitude=1.9pt, segment length=2.9pt,aspect=0.9pt}, line width=0.6pt, color=internationalorange] (-10:18pt) -- (28pt,7pt); 
  \draw[-, line width=0.8pt, color=GWmerger] (d) -- (e) -- (f);
  \draw[xshift=0pt, -, line width=0.8pt, color=GWmerger] (d) -- (e) node[currarrow, pos=0.5, xscale=-2.5, sloped, scale=0.25] {};
  \draw[xshift=0pt, -, line width=0.8pt, color=GWmerger] (e) -- (f) node[currarrow, pos=0.5, xscale=-2.5, sloped, scale=0.25] {};
  \draw[line width=0.8pt] (-10:8pt) -- (-10:43pt); 
\end{tikzpicture}
\end{aligned}
=  {-1 \over q_{12}^2}\, \mathcal{S}_i^\mu(q_{12})\,\bar{u}_{\lambda}(q_1)\,\gamma_{\mu}\,v_{\lambda'}(q_2)\, \mathbf{T}_i^{a}\,t^a_{j\bar{\imath}},
\end{align}
where $q_{12} = q_1+ q_2$, $q_1$ and $q_2$ denote the momenta of the quark and antiquark, $\lambda$ and $\lambda'$ denote the spins of the quark and antiquark, and $j$ and $\bar{\imath}$ are fundamental and anti-fundamental indices of the Lie algebra $\operatorname{su}(N_c)$ carried by the quark and antiquark respectively.
Like the soft-gluon case, the full current for the emission of a massless quark pair can be obtained by summing over external hard lines \cite{Catani:1999ss}, i.e.,
\begin{align}\label{J-soft-qq}
\mathbf{J}_{q\bar{q}}^{j\,\bar{\imath}}(q_1,q_2) 
= \sum_i {-1 \over q_{12}^2}\, \mathcal{S}_i^\mu(q_{12})\,\gamma_{\mu}\, \mathbf{T}_i^{a}\,t^a_{j\bar{\imath}}
= {-1 \over q_{12}^2}  \mathbf{J}^{a;\mu}_g(q_{12})\,\gamma_{\mu}\,t_{j\bar{\imath}}^a,
\end{align}
or in the vector form in color-spin space
\begin{align}\label{J-soft-qq-vec}
    \mathbf{J}_{q\bar{q}}(q_1,q_2) = \bra{j\,\bar{\imath}} \otimes \bra{\lambda\lambda'}
    \,\bar{u}_{\lambda}(q_1)\,\mathbf{J}_{q\bar{q}}^{j\,\bar{\imath}}(q_1,q_2)\,v_{\lambda'}(q_2)\, 
\end{align}
The last equality in \eqref{J-soft-qq} gives the relation between the currents for a soft quark pair of a single-gluon emissions, which reflects the fact that the quark-antiquark are produced from a off-shell soft gluon line.

\vskip 5pt
In the soft limit, the factorization property of scattering amplitudes naturally induces factorization behavior of the squared matrix elements.
Simply using formula \eqref{amp-vec-sq} and the current \eqref{J-single-gluon-vec}, one can immediately find
\begin{align}\label{J-single-gluon-sq}
\mathscr{S}_{q}\, \big|\mathcal{M}_{gf_1\cdots f_n}\big|^2
=  - (\mu^{\epsilon}g_s)^{2}   \sum_{i,j=1}^n \mathcal{S}_{ij}(q)\, \big|\mathcal{M}_{f_1\cdots f_n}^{(ij)} \big|^2,
\end{align}
with
\begin{align}\label{dipole-corr-amp-sq}
\big|\mathcal{M}_{f_1\cdots f_n}^{(ij)} \big|^2 \equiv \Bra{\mathcal{M}_{f_1 \cdots f_n}}  \mathbf{T}_i\cdot \mathbf{T}_j \Ket{\mathcal{M}_{f_1 \cdots f_n} }.
\end{align}
Below we explain the notations appearing in the above formula.
The well-known eikonal function is given by
\begin{align}\label{def-eikonal-function}
    \mathcal{S}_{ij}(q) = -d_{\mu\nu}(q,r)\, \mathcal{S}_i^\mu(q) \mathcal{S}_j^\nu(q)
    \simeq {p_i\cdot p_j  \over  (q\cdot p_i)( q\cdot p_j)}\,,
\end{align}
with $d_{\mu\nu}$ the tensor from the completeness relation of polarizations, i.e.,
\begin{align}
d_{\mu\nu}(q,r) = \sum_{\lambda=1}^{D-2} \epsilon^{\lambda}_{\mu}(q,r) \big[\epsilon^{\lambda}_{\nu}(q,r)\big]^\ast = -g_{\mu\nu}+\frac{q_{\mu} r_\nu+r_\mu q_{\nu}}{q\cdot r}\,.
\end{align}
where the summation is carried over $D{-}2$ physical degrees of freedom in the axial gauge.
It is also noteworthy that one dropped the terms dependent on reference vector $r$ since they do not contribute the squared matrix element in \eqref{J-single-gluon-sq} due the colour conservation illustrated in \eqref{color-conservation}.
The color operator is defined as
\begin{align}\label{def-dipole}
\mathbf{T}_i\cdot \mathbf{T}_j =  \mathbf{T}_j\cdot \mathbf{T}_i =  \mathbf{T}_i^a \mathbf{T}_j^a
\end{align}
which generates a {\it dipole}-type color correlation in the squared amplitude \eqref{dipole-corr-amp-sq}.
According to formulas \eqref{def-eikonal-function} and \eqref{def-dipole}, an obvious yet important property is that the amplitude is symmetric under the exchange of $i$ and $j$, referred to as the dipole symmetry.

\vskip 5pt
Now let us turn to consider the square of the current of soft quark-antiquark pair in \eqref{J-soft-qq} or \eqref{J-soft-qq-vec}.
Directly from \eqref{J-soft-qq-vec}, one obtains
\begin{align}\label{J-soft-qq-sq}
\mathscr{S}_{q_1q_2}  \big| \mathcal{M}_{q \bar{q} f_1\cdots f_n} \big|^2
= (\mu^{2\epsilon}g_s^2)^2\,T_F\,\sum_{i,j=1}^n \mathcal{I}_{ij}(q_1,\!q_2)\,
\big| \mathcal{M}_{f_1\ldots f_n}^{(ij)} \big|^2 \,,
\end{align}
where $T_F=\frac{1}{2}$ is explained in \eqref{Casimir-operator}, and the kinematic function is defined as
\begin{align}\label{eikonal-function-qq}
\mathcal{I}_{ij}(q_1,q_2) 
&=  {1 \over q_1\cdotnew q_2} \mathcal{S}_i^{\mu}(q_{12})\, \mathcal{S}_j^{\nu}(q_{12})\, d_{\mu\nu}(q_1,q_2) 
\\[0.3 em]
&= { (p_i\cdot q_1) (p_j\cdot q_2) + (p_i\cdot q_2) (p_j\cdot q_1) - (q_1\cdot q_2) (p_i\cdot p_j)
\over (q_1 \cdot q_2)^2 (p_i\cdot q_{12}) (p_j\cdot q_{12})}.
\end{align}
Let us make some essential observations at this point.
First, like the single gluon case, the squared current for the soft-quark emission in \eqref{J-soft-qq-sq} also owns the dipole symmetry under exchange of the emitters $(i\leftrightarrow j)$. Additionally, the kinematic function $\mathcal{I}_{ij}$ given in \eqref{eikonal-function-qq} features charge conjugation symmetry under the exchange of the quark and antiquark labels
\begin{align}\label{}
\mathcal{I}_{ij}(q_1,q_2) = \mathcal{I}_{ij}(q_2,q_1) \,.
\end{align}

To summarize, we briefly review the eikonal formalism of tree-level currents for multiple soft parton emission from a hard QCD scattering.
We also show two most representative examples, a single-soft gluon and a double soft quark-antiquark pair, to illustrate how the formalism works for constructing currents in detail.

\section{The tree-level current for two pairs of soft quarks}
\label{sec-soft-qqqq-current}

We present the core result of this study in this section. We first construct the tree-level current for the emission of two soft quark-antiquark pairs following the eikonal formalism outlined above. We then proceed to compute the squared current and analyze the resulting color correlations.

\subsection{The soft current}
\begin{figure}[H]
\centering
\includegraphics[width=153mm]{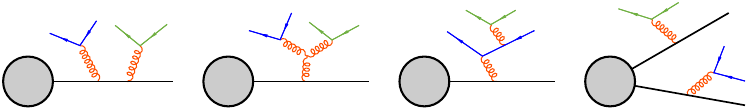}
\caption{Diagrams contributing to the tree-level current for the emission of two soft quark-antiquark pairs. Red and green lines represent the soft pairs $q\bar{q}$ and $Q\bar{Q}$, respectively. For the contributions from the first and third diagrams, we should also include the contributions from the exchange of two soft-quark pairs.}
\label{diagrams-soft-qqqq}
\end{figure}

Without loss of generality, we assume that the two quark-antiquark pairs, denoted as $q\bar{q} Q\bar{Q}$, carry momenta $\{q_1,q_2,q_3,q_4\}$ and color indices $\{i,\bar{\imath}, j,\bar{\jmath}\}$, as well as helicities $\{\lambda_1,\lambda_2,\lambda_3,\lambda_4\}$.
Recalling that a soft quark-antiquark pair can be produced from an intermediate soft gluon.
According to the eikonal Feynman rules, we need to consider the four types of diagrams displayed in Figure\,\ref{diagrams-soft-qqqq}.

Let us initiate our analysis with the fourth diagram in Figure\,\ref{diagrams-soft-qqqq}, which yields
\begin{align}\label{J-qqQQ-fig-4}
{\big[ \bar{u}_1 \gamma_\mu\, v_2 \big] \big[ \bar{u}_3 \gamma_\nu\, v_4 \big]  
t^a_{i\bar{\imath}} t^b_{j\bar{\jmath}}
\over  q_{12}^2 q_{34}^2}
\sum_{k\ne l}
{\cal S}_{k}^{\mu}(q_{12})\,{\cal S}_{l}^{\nu}(q_{34})\,
\mathbf{T}_k^{a}\,\mathbf{T}_l^{b},
\end{align}
where, for notational compactness, we suppress the momentum and spin-index dependence on the spinors:
\begin{align}\label{omit-spin-on-spinors}
\bar{u}_1 \equiv \bar{u}_{\lambda_1}(q_1), \quad
v_2 \equiv v_{\lambda_2}(q_2), \quad
\bar{u}_3 \equiv \bar{u}_{\lambda_3}(q_3), \quad
v_4 \equiv v_{\lambda_4}(q_4).
\end{align}
Next, the first diagram in Figure\,\ref{diagrams-soft-qqqq} contributes as follows:
\begin{align}\label{J-qqQQ-fig-1-a}
{\big[ \bar{u}_1 \gamma_\mu v_2 \big] \big[ \bar{u}_3 \gamma_\nu v_4 \big]  
t^a_{i\bar{\imath}} t^b_{j\bar{\jmath}}
\over  q_{12}^2 q_{34}^2}
\sum_{k=1}^n
&
{\cal S}_{k}^{\mu}(q_{1234}) \Big[
{\cal S}_{k}^{\nu}(q_{12})\,\mathbf{T}_k^{a}\mathbf{T}_k^{b}
+ {\cal S}_{k}^{\nu}(q_{34})\,\mathbf{T}_k^{b}\mathbf{T}_k^{a}
\Big]
\end{align}
It is useful to symmetrize the colour charges therough the following identity
\begin{align}\label{}
  \mathbf{T}_i^{a} \mathbf{T}_i^{b} \,=\, {1\over 2}\{\mathbf{T}_i^{a}, \mathbf{T}_i^{b}\}
  + {1\over 2}[\mathbf{T}_i^{a}, \mathbf{T}_i^{b}]
  \,=\, {1\over 2}\{\mathbf{T}_i^{a}, \mathbf{T}_i^{b}\}
  + {i \over 2} f^{abc}\mathbf{T}_i^{c}.
\end{align}
This allows us to decompose \eqref{J-qqQQ-fig-1-a} into symmetric part and a anti-symmetric part in two color charges
\begin{align}\label{}
&
{\big[ \bar{u}_1 \gamma_\mu v_2 \big] \big[ \bar{u}_3 \gamma_\nu v_4 \big]  
t^a_{i\bar{\imath}} t^b_{j\bar{\jmath}}  \over  q_{12}^2 q_{34}^2}
\nonumber\\
\label{J-qqQQ-fig-1-b}
&\quad
\times \sum_{k=1}^n
\bigg(
{1\over 2}\{\mathbf{T}_k^{a}, \mathbf{T}_k^{b}\}\,{\cal S}_{k}^{\mu}(q_{12}){\cal S}_{k}^{\nu}(q_{34})
+ {i \over 2}f^{abc}\mathbf{T}_k^{c}
{\cal S}_{k}^{\mu}(q_{1234}) \big[{\cal S}_{k}^{\nu}(q_{12}) - {\cal S}_{k}^{\nu}(q_{34})\big]
\bigg).
\end{align}

Combining the symmetric part in \eqref{J-qqQQ-fig-1-b} with \eqref{J-qqQQ-fig-4} yields
\begin{align}\label{}
\mathbf{A}_1 = 
{\big[ \bar{u}(q_1) \gamma_\mu\, v(q_2) \big] \big[ \bar{u}(q_3) \gamma_\nu\, v(q_4) \big]  
\over  q_{12}^2 q_{34}^2}
\times {1 \over 2} \big\{\mathbf{J}_g^{a;\mu}(q_{12}), \,\mathbf{J}_g^{b;\nu}(q_{34})\big\}
t^a_{i\bar{\imath}} t^b_{j\bar{\jmath}} \,,
\label{J-qqQQ-A1}
\end{align}
while combining the anti-symmetric part in \eqref{J-qqQQ-fig-1-b} and the contributions from the second diagram in Figure\,\ref{diagrams-soft-qqqq} gives
\begin{align}
\mathbf{N} =\,
&
{\big[ \bar{u}_1 \gamma_\mu v_2 \big] \big[ \bar{u}_3 \gamma_\nu v_4 \big]  
t^a_{i\bar{\imath}} t^b_{j\bar{\jmath}}  \over  q_{12}^2 q_{34}^2}
\nonumber\\[0 em]
&
~ \times\sum_{i=1}^n
if^{abc}\mathbf{T}_i^{c}\, {\cal S}_{i}^{\alpha}(q_{1234}) 
\bigg(
{1 \over 2}\delta^{\mu}_\alpha  \big[{\cal S}_{i}^{\nu}(q_{12}) - {\cal S}_{i}^{\nu}(q_{34})\big] 
- {g_{\alpha\sigma} \over  q_{1234}^2} V^{\mu\nu\sigma}(q_{12},q_{34})
\bigg),
\label{J-qqQQ-N}
\end{align}
with
\begin{align}\label{}
V^{\mu\nu\alpha}(q_{12},q_{34}) \equiv 
g^{\mu\nu} (q_{12} {-} q_{34})^{\alpha}
+ g^{\nu\alpha} (q_{12} {+} 2q_{34})^{\mu}
- g^{\mu\alpha} (q_{34} {+} 2q_{12})^{\nu}.
\end{align}
Lastly, the contribution from the third diagram in Figure\,\ref{diagrams-soft-qqqq} is given by
\begin{align}\label{J-qqQQ-A2}
\mathbf{A}_2 =\,&
{[ \bar{u}_3 \gamma^\nu v_4 ] t^b_{j\bar{\jmath}} \over  q_{1234}^2 q_{34}^2}
\sum_{k=1}^n  \mathbf{T}_k^{a} {\cal S}_{k}^{\mu}(q_{1234})\,
\bar{u}_1\bigg[
{(t^b t^a)_{i\bar{\imath}} \over q_{134}^2} \gamma_\nu \slashed{q}_{134}\gamma_\mu 
- {(t^a t^b)_{i\bar{\imath}} \over q_{234}^2} \gamma_\mu \slashed{q}_{234}\gamma_\nu 
\bigg]
v_2
\nonumber\\[0.2 em]
&
+ (q\bar{q}\leftrightarrow Q\bar{Q}),
\end{align}
where the notation $(q\bar{q}\leftrightarrow Q\bar{Q})$ represents a operation where we exchange all quantum numbers for the two soft pairs, including spins, color indices and  momenta. 

\vskip 5pt
Collecting all contributions, the complete tree-level current for the emission of two soft quark pairs is given by
\begin{align}\label{J-soft-qqQQ}
\mathbf{J}_{q\bar{q}Q\bar{Q}}^{i\bar{\imath}  j\bar{\jmath}; \lambda_1 \cdots \lambda_4}(q_1,q_2,q_3,q_4) 
= \mathbf{A}_1 + \mathbf{A}_2 + \mathbf{N},
\end{align}
or in the vector form in color-spin space
\begin{align}\label{J-soft-qqQQ-vec}
\mathbf{J}_{q\bar{q}Q\bar{Q}}(q_1,q_2,q_3,q_4) = \bra{i\,\bar{\imath}\, j\,\bar{\jmath}} \otimes \bra{\lambda_1\lambda_2\lambda_3\lambda_4}
\mathbf{J}_{q\bar{q}Q\bar{Q}}^{i\bar{\imath}  j\bar{\jmath}; \lambda_1 \cdots \lambda_4}(q_1,q_2,q_3,q_4)\, .
\end{align}
Note that we have omitted the dependence on helicity indices in $\mathbf{A}_i$ and $\mathbf{N}$, as indicated in \eqref{omit-spin-on-spinors}.

\vskip 5pt
Now, we highlight several key properties of this current. First, the term $\mathbf{A}_1$ owns a factorization structure in which it factorizes into a (off-shell) double soft gluon current contracted with the two quark-antiquark currents. Second, as expected, in addition to $\mathbf{A}_2$ in \eqref{J-soft-qqQQ}, the full current is also symmetric under the exchange of two quark pairs, i.e.~$q\bar{q} \leftrightarrow Q\bar{Q}$. Finally, term $\mathbf{N}$ in \eqref{J-qqQQ-N} is pure non-Abelian, whereas the other two terms can be reduced to the  corresponding current in Abelian quantum electrodynamics (QED) by simply treating color charges as electrical charges and setting the fundamental representation generators to unity.

\subsection{The squared current}

Next, we compute the square of the current obtained previously.
The full result for the squared current has a similar color-correlation structure with the soft emission of a quark pair in association with a gluon discussed in \cite{DelDuca:2022noh}
\begin{align}\label{J-qqQQ-sq}
\mathscr{S}_{q_1\cdots q_4} \big|\mathcal{M}_{q\bar{q}Q\bar{Q} f_1\cdots f_n}\big|^2
=  \big(&\mu^{2\epsilon} g_s^2\big)^{4}
\Bigg[
{T_F^2 \over 2} \sum_{i,j,k,l=1}^n \mathcal{Q}_{ijkl} \big|\mathcal{M}_{f_1\cdots f_n}^{(ij;kl)} \big|^2
\nonumber\\
&
+ T_F\sum_{i,j,k=1}^n \mathcal{Q}_{ijk} \big|\mathcal{M}_{f_1\cdots f_n}^{(ijk)} \big|^2
+ T_F^2\sum_{i,j=1}^n \mathcal{Q}_{ij} \big|\mathcal{M}_{f_1\cdots f_n}^{(ij)} \big|^2
\Bigg],
\end{align}
where the dipole-correlated squared amplitude is defined in \eqref{dipole-corr-amp-sq} while the symmetrized doule-dipole-correlated and the triple-color correlated squared amplitudes are defined as \cite{DelDuca:2022noh}
\begin{align}\label{tow-dipole-corr-amp-sq}
\big|\mathcal{M}_{f_1\cdots f_n}^{(ij;kl)} \big|^2 
\equiv \Bra{\mathcal{M}_{f_1 \cdots f_n}}  \{\mathbf{T}_i\cdot \mathbf{T}_j, \mathbf{T}_k\cdot \mathbf{T}_l \} \Ket{\mathcal{M}_{f_1 \cdots f_n} }
\end{align}
and
\begin{align}\label{triple-dipole-corr-amp-sq}
\big|\mathcal{M}_{f_1\cdots f_n}^{(ijk)} \big|^2 
\equiv d^{abc} \Bra{\mathcal{M}_{f_1 \cdots f_n}}  \mathbf{T}_i^a \mathbf{T}_j^b \mathbf{T}_k^c \Ket{\mathcal{M}_{f_1 \cdots f_n} }
\end{align}
with the symmetric structure constant
\begin{align}\label{def-d-abc}
d^{abc} \equiv  2 \operatorname{Tr}\big[\{t^a, t^b\} t^c\big].
\end{align}

We now proceed to elaborate on the derivation details on \eqref{J-qqQQ-sq} and provide the explicit expressions for various kinematical coefficients. 
First, the derivation for the square of $\mathbf{A}_1$ is quite similar with the derivation for the squared current of double-soft gluon emission in \cite{Czakon:2011ve}
\begin{align}\label{J-qqQQ-sq-A1}
\big|\mathbf{A}_1 \big|^2 
&=
{T_F^2 \over 4} \sum_{i,j,k,l} 
\big\{\mathbf{T}_{i}^{a},\mathbf{T}_{j}^{b}\big\} \big\{\mathbf{T}_{k}^{a},\mathbf{T}_{l}^{b}\big\}\,
\mathcal{Q}_{ijkl}
\nonumber\\
&=
{T_F^2 \over 16} \sum_{i,j,k,l} 
\Big(
\big\{ \big\{\mathbf{T}_{i}^{a},\mathbf{T}_{j}^{b}\big\}, \big\{\mathbf{T}_{k}^{a},\mathbf{T}_{l}^{b}\big\}  \big\}
+ \big\{ \big\{\mathbf{T}_{i}^{a},\mathbf{T}_{j}^{b}\big\}, \big\{\mathbf{T}_{k}^{a},\mathbf{T}_{l}^{b}\big\}  \big\}
\Big)
\mathcal{Q}_{ijkl},
\end{align}
with
\begin{align}\label{def-Qijkl}
\mathcal{Q}_{ijkl}(q_1,q_2,q_3,q_4) = \mathcal{I}_{i k}(q_1,q_2)\, \mathcal{I}_{j l}(q_3,q_4),
\end{align}
where the eikonal function $\mathcal{I}_{ij}$ is defined in \eqref{eikonal-function-qq} and we used the symmetry $\mathcal{I}_{ij} = \mathcal{I}_{ji}$.
Then noticing Czakon's identity \cite{Czakon:2011ve}
\begin{align}\label{Czakon-id-1}
\big\{\{ \mathbf{T}_{i}^{a}, \mathbf{T}_{j}^{b}\}, \{\mathbf{T}_{k}^{a}, \mathbf{T}_{l}^{b}\} \big\}
+ \big\{\{\mathbf{T}_{i}^{a}, & \mathbf{T}_{l}^{b}\},\{\mathbf{T}_{k}^{a}, \mathbf{T}_{j}^{b}\} \big\}
\nonumber\\[0.3 em]
= 8\big\{\mathbf{T}_{i} \!\cdot\! \mathbf{T}_{k},\, \mathbf{T}_{j} \!\cdot\! \mathbf{T}_{l} \big\}
+2 C_{A} \Big[
&
3 \delta_{i l} \delta_{j k}\, (\mathbf{T}_{i} \!\cdot\! \mathbf{T}_{j})
+ 3\delta_{i j} \delta_{k l}\, (\mathbf{T}_{i} \!\cdot\! \mathbf{T}_{k})
-2 \delta_{i j} \delta_{j k}\, (\mathbf{T}_{i} \!\cdot\! \mathbf{T}_{l})
\nonumber\\
&
-2 \delta_{i j} \delta_{j l}\, (\mathbf{T}_{i} \!\cdot\! \mathbf{T}_{k})
-2(\delta_{i k} \delta_{k l}+\delta_{j k} \delta_{k l}) (\mathbf{T}_{i} \!\cdot\! \mathbf{T}_{j})
\Big],
\end{align}
we can immediately reduce the expression in \eqref{J-qqQQ-sq-A1} into a double-dipole part and a dipole part.
Precisely,
\begin{align}
\big|\mathbf{A}_1 \big|^2 
=~
&{T_F^2 \over 2} \sum_{i,j,k,l} 
\big\{\mathbf{T}_{i} \!\cdot\! \mathbf{T}_{k},\, \mathbf{T}_{j} \!\cdot\! \mathbf{T}_{l} \big\}
\mathcal{Q}_{i j k l}
\nonumber\\
&+
{C_A T_F^2 \over 4} \sum_{i,j} 
\big(\mathbf{T}_{i} \!\cdot\! \mathbf{T}_{j}\big)
\big(
3\mathcal{Q}_{i i j j}
-3\mathcal{Q}_{i i i j}
- \mathcal{Q}_{i j j j}
\big).
\label{J-qqQQ-sq-A1-dipole-final}
\end{align}

Next we consider the product between $\mathbf{A}_1$ and $\mathbf{N}$
\begin{align}
\mathbf{A}_1^\dagger\mathbf{N} + \mathbf{N}^\dagger \mathbf{A}_1
&=
T_F^2 \sum_{j,k,l} 
{-i \over 2} f^{abc}\big[ \big\{\mathbf{T}_{j}^{a}, \mathbf{T}_{k}^{b}\big\}, \mathbf{T}_{l}^{c}\big]\, \mathcal{I}_{jkl}
\nonumber\\
&=  T_F^2 C_A 
\sum_{i,j} \big( \mathcal{I}_{ijj} - \mathcal{I}_{iji} \big)\, (\mathbf{T}_{i} \!\cdot\! \mathbf{T}_{j}),
\label{J-qqQQ-sq-A1-N}
\end{align}
with
\begin{align}\label{}
\mathcal{I}_{ijl}(q_1,q_2,q_3,q_4) \equiv
\,&{4 \over q_{12}^2 q_{34}^2}
d_{\mu_1\mu_2}(q_1,q_2) d_{\nu_1\nu_2}(q_3,q_4)
{\cal S}_{i}^{\mu_1}(q_{12})\,{\cal S}_{j}^{\nu_1}(q_{34})
\\
\nonumber
&\times
{\cal S}_{l}^{\alpha}(q_{1234}) \bigg(
{1 \over 2}\delta^{\mu_2}_\alpha  \big[{\cal S}_{l}^{\nu_2}(q_{12}) - {\cal S}_{l}^{\nu_2}(q_{34})\big] 
- {g_{\alpha\sigma} \over  q_{1234}^2} V^{\mu_2\nu_2\sigma}(q_{12},q_{34})
\bigg).
\end{align}
To get \eqref{J-qqQQ-sq-A1-N} we used the identity \cite{Czakon:2011ve}
\begin{align}\label{Czakon-id-2}
i f^{a b c}\big[ \{\mathbf{T}_{i}^{a}, \mathbf{T}_{j}^{b} \}, \mathbf{T}_{k}^{c}\big] 
= 2 C_A (\mathbf{T}_{i} \cdotnew \mathbf{T}_{j}) (\delta_{i k} - \delta_{j k}).
\end{align}
We see that this term is fully reduced to dipole correlation.

\vskip 5pt
A simple observation is that, in the product between $\mathbf{A}_1$ and $\mathbf{A}_2$, we will see the trace of three fundamental representation generators $\tr(t^a t^b t^c)$, for which we decompose it into a full symmetric part and a full anti-symmetric part, i.e.,
\begin{align}\label{}
\operatorname{tr}\big(t^{a} t^{b} t^{c}\big) = \frac{1}{4}d^{a b c} + \frac{i}{2}\,T_F\, f^{a b c}.
\end{align}
Then the product of $\mathbf{A}_1$ and $\mathbf{A}_2$ is finally given by
\begin{align}
\mathbf{A}_1^\dagger\mathbf{A}_2 + \mathbf{A}_2^\dagger\mathbf{A}_1
&=
T_F\sum_{i,j,l}
\bigg[
 {i \over 2} f^{abc} T_F \big[\big\{\mathbf{T}_i^{a}, \mathbf{T}_j^{b}\big\},\, \mathbf{T}_l^{c} \big]\mathcal{K}_{ijl}
+  {1 \over 4} d^{abc} \big\{\big\{\mathbf{T}_i^{a}, \mathbf{T}_j^{b}\big\},\, \mathbf{T}_l^{c} \big\}
\mathcal{Q}_{ijl}
\bigg]
\nonumber\\
&=
C_A T_F^2 \sum_{i,j}
(\mathbf{T}_i \cdotnew \mathbf{T}_j) (\mathcal{K}_{iji} - \mathcal{K}_{ijj})
+
T_F\sum_{i,j,l} \mathcal{Q}_{ijl}\,
d^{abc} \mathbf{T}_i^{a} \mathbf{T}_j^{b} \mathbf{T}_l^{c},
\label{J-qqQQ-sq-A1*A2}
\end{align}
where
\begin{align}\label{def-Qijl}
\mathcal{K}_{ijl} &= \mathcal{E}_{ijl} + \mathcal{F}_{ijl} - \mathcal{G}_{ijl} - \mathcal{H}_{ijl},
\qquad
\mathcal{Q}_{ijl} = \mathcal{E}_{ijl} - \mathcal{F}_{ijl} + \mathcal{G}_{ijl} - \mathcal{H}_{ijl},
\end{align}
with
\begin{align}
\mathcal{E}_{ijl} &\equiv  
{\tr(\gamma_{\mu} \slashed{q}_1 \gamma^{\sigma} \slashed{q}_{134}\gamma_{\alpha} \slashed{q}_2)\, d_{\nu\sigma}(q_3,q_4)
 \over  q_{12}^2 q_{34}^2 q_{1234}^2 q_{134}^2} 
{\cal S}_{i}^{\mu}(q_{12})\,{\cal S}_{j}^{\nu}(q_{34})
{\cal S}_{l}^{\alpha}(q_{1234})\,,
\\[0.5 em]
\mathcal{F}_{ijl} &\equiv  
{\tr(\gamma_{\mu} \slashed{q}_1 \gamma_{\alpha} \slashed{q}_{234}\gamma^{\sigma} \slashed{q}_2)\, d_{\nu\sigma}(q_3,q_4)
 \over  q_{12}^2 q_{34}^2  q_{1234}^2 q_{234}^2}  
{\cal S}_{i}^{\mu}(q_{12})\,{\cal S}_{j}^{\nu}(q_{34})
{\cal S}_{l}^{\alpha}(q_{1234})\,,
\\[0.5 em]
\mathcal{G}_{ijl} &\equiv  
{\tr(\gamma_{\nu} \slashed{q}_3 \gamma^{\sigma} \slashed{q}_{123}\gamma_{\alpha}  \slashed{q}_4)\, d_{\mu\sigma}(q_1,q_2)
 \over  q_{12}^2 q_{34}^2  q_{1234}^2 q_{123}^2} 
{\cal S}_{i}^{\mu}(q_{12})\,{\cal S}_{j}^{\nu}(q_{34})
{\cal S}_{l}^{\alpha}(q_{1234})\,,
\\[0.5 em]
\mathcal{H}_{ijl} &\equiv  
{\tr(\gamma_{\nu} \slashed{q}_3 \gamma_{\alpha} \slashed{q}_{124}\gamma^{\sigma} \slashed{q}_4)\, d_{\mu\sigma}(q_1,q_2)
 \over  q_{12}^2 q_{34}^2  q_{1234}^2 q_{124}^2} 
{\cal S}_{i}^{\mu}(q_{12})\,{\cal S}_{j}^{\nu}(q_{34})
{\cal S}_{l}^{\alpha}(q_{1234})\,.
\end{align}
As expected, in \eqref{J-qqQQ-sq-A1*A2} the term linear in $f^{abc}$ can ultimately be simplified to a dipole correlation using Czakon's identity \eqref{Czakon-id-2}, while the $d^{abc}$ part yields a triple-parton color correlation, similar to the emission of a soft quark-antiquark pair in association with a gluon.

\vskip 5pt
Lastly, we turn our attention to $|\mathbf{A}_2+\mathbf{N}|^2$, which, as expected, contributes only to the dipole correlation.
More specifically, we find
\begin{align}
|\mathbf{N}|^2 &\sim f^{abc}f^{abe}\, \mathbf{T}_i^{c} \mathbf{T}_j^{e} = C_A (\mathbf{T}_i\cdotnew \mathbf{T}_j)\,,
\\
\mathbf{N}^\dagger \mathbf{A}_2 &\sim f^{abc} \tr (t^{a} t^b t^e)\, \mathbf{T}_i^{c} \mathbf{T}_j^{e} = {i \over 2} T_FC_A (\mathbf{T}_i\cdotnew \mathbf{T}_j)\,,
\end{align}
and in $|\mathbf{A}_2|^2$ there are the following three types of terms
\begin{align}
\tr (t^{a} t^{a} t^b t^c)\, \mathbf{T}_i^{b} \mathbf{T}_j^{c} &= T_FC_F (\mathbf{T}_i\cdotnew \mathbf{T}_j),
\\[0.35 em]
\tr (t^{a} t^{b} t^a t^c)\, \mathbf{T}_i^{b} \mathbf{T}_j^{c} &= T_F \big(C_F - \tfrac{1}{2} C_A\big) (\mathbf{T}_i\cdotnew \mathbf{T}_j),
\\[0.35 em]
\tr (t^{a} t^b t^c)\, \tr (t^{a} t^{b} t^{e}) \mathbf{T}_i^{c} \mathbf{T}_j^{e} &= 
T_F^2 \big( 2 C_F - \tfrac{3}{4} C_A  - \tfrac{1}{4} C_A  \big) (\mathbf{T}_i\cdotnew \mathbf{T}_j).
\end{align}
With these identities, the remaining steps in the derivation involve primarily extensive but straightforward Lorentz contraction algebra. While we omit further derivation details, the final result takes the following form:
\begin{align}\label{J-qqQQ-sq-A2+N}
|\mathbf{A}_2 + \mathbf{N}|^2  
=  \sum_{i,j=1}^{n} T_F^2 \Big[ \mathcal{R}_{ij}^\text{(ab)} C_A  + \mathcal{R}_{ij}^\text{(nab)} C_F \Big] (\mathbf{T}_i\cdotnew \mathbf{T}_j).
\end{align}
The explicit expressions for $\mathcal{R}_{ij}^\text{(ab)}$ and $\mathcal{R}_{ij}^\text{(nab)}$ are lengthy, and we include them in an computer-readable ancillary file.

\vskip 5pt
We now compile all previously-derived terms to provide full analytic expressions for the three kinematical functions in the squared amplitude \eqref{J-qqQQ-sq}. The functions $\mathcal{Q}_{ijkl}$ and $\mathcal{Q}_{ijl}$ are given by equations \eqref{def-Qijkl} and \eqref{def-Qijl} while the kinematical coefficient $\mathcal{Q}_{ij}$ receives the contributions from \eqref{J-qqQQ-sq-A1*A2}, \eqref{J-qqQQ-sq-A1-N}, \eqref{J-qqQQ-sq-A1-dipole-final}, \eqref{J-qqQQ-sq-A2+N}.
Their combination yields:
\begin{align}\label{}
\mathcal{Q}_{ij} = 
C_F\mathcal{R}_{ij}^\text{(nab)}
+ C_A \Big[
\mathcal{R}_{ij}^\text{(ab)}
&+ {1 \over 4}
(
3\mathcal{Q}_{i i j j}
-3\mathcal{Q}_{i i i j}
- \mathcal{Q}_{i j j j}
)
\nonumber\\
&+  \mathcal{I}_{ijj} - \mathcal{I}_{iji}
+ \mathcal{K}_{iji} - \mathcal{K}_{ijj}
\Big].
\label{final-Qij}
\end{align}

A few observations about the squared current are in order.
First, $\mathcal{Q}_{ijkl}$ in \eqref{def-Qijkl} is manifestly symmetric under the exchange of indices $(i\leftrightarrow k)$ or $(j\leftrightarrow l)$ as well as under the exchange of momenta $(q_1\leftrightarrow q_2)$ or $(q_3\leftrightarrow q_4)$, due to the symmetry properties of the eikonal function $\mathcal{I}_{ik}$. Second, while the triple function $\mathcal{Q}_{ijl}$ defined in \eqref{def-Qijl} is not symmetric in hard parton labels, only its fully symmetric component contributes, as the three-parton correlated squared matrix element defined in \eqref{triple-dipole-corr-amp-sq} is fully symmetric. Like $\mathcal{Q}_{ijkl}$, $\mathcal{Q}_{ijl}$ is independent of the masses of the hard particles. Similarly, although the dipole function $\mathcal{Q}_{ij}$ in \eqref{final-Qij} does not inherently display symmetry in hard parton labels, only its symmetric part contributes because the dipole operator $(\mathbf{T}_i\cdotnew \mathbf{T}_j)$ is symmetric. Moreover, the coefficient of $C_F$ in $\mathcal{Q}_{ij}$ is independent of masses, while the coefficient of $C_A$ in $\mathcal{Q}_{ij}$ linearly depends on $m_i^2$ and $m_j^2$.

\subsection{Strongly-ordered soft limit}

We now consider a strongly-ordered soft limit, where one pair of quarks is significantly softer than the other. Concretely, we analyze the sub kinematical regime in which the momenta of one pair are considerably smaller than those of the other pair. Without loss of generality, we examine the soft current in the following limit: 
\begin{align}\label{Strongly-ordered-limit-qqQQ}
E_1, E_2 \gg E_3, E_4.
\end{align}
In this limit, we find that the double-dipole function remains exact:
\begin{align*}
&\mathcal{Q}_{ijkl} \rightarrow \mathcal{Q}_{ijkl} = \mathcal{I}_{i k}(q_1,q_2)\, \mathcal{I}_{j l}(q_3,q_4),
\end{align*}
while the other functions, $\mathcal{Q}_{ijl}$ and $\mathcal{Q}_{ij}$, are significantly simplified
\begin{align}
&\mathcal{Q}_{ijl} \rightarrow
{2d_{\mu\nu}(q_1,q_2)\, d_{\alpha\sigma}(q_3,q_4)\, {\cal S}_{i}^{\mu}(q_{12}) {\cal S}_{l}^{\nu}(q_{12}) {\cal S}_{j}^{\alpha}(q_{34})  \over  q_{12}^2 q_{34}^2}
\bigg[
{q_1^\sigma   \over   q_{1}\cdotnew q_{34} } 
 -
{q_2^{\sigma}  \over  q_2\cdotnew q_{34} }  
\bigg],
\\[0.35 em]
&\mathcal{Q}_{ij} \rightarrow
{d_{\mu\nu}(q_1,q_2)\, d_{\alpha\sigma}(q_3,q_4)\, \mathcal{S}_i^{\mu}(q_{12})   \over  2q_{12}^2 q_{34}^2}
\Bigg\{
C_A \bigg[
 \frac{4\mathcal{S}_j^{\nu }(q_{12})\, q_1^\alpha q_2^\sigma}{(q_1\cdotnew q_{34}) (q_2\cdotnew q_{34})}
\nonumber\\
&
\qquad\qquad
+ 2\Big(\mathcal{S}_i^{\nu }(q_{12}) \mathcal{S}_j^{\alpha }(q_{34})  - [\mathcal{S}_i^{\alpha }(q_{34}) + \mathcal{S}_j^{\alpha }(q_{34})] \mathcal{S}_j^{\nu }(q_{12}) \Big)
\bigg( \frac{q_1^\sigma }{q_1\cdotnew q_{34}} +   \frac{q_2^\sigma }{q_2\cdotnew q_{34}} \bigg)
\nonumber\\
&
\qquad\qquad
+ \mathcal{S}_j^{\sigma }(q_{34}) \Big(4 \mathcal{S}_i^{\alpha }(q_{34}) S_j^{\nu }(q_{12}) 
- \mathcal{S}_i^{\alpha }(q_{34}) \mathcal{S}_i^{\nu }(q_{12})    
- 3 \mathcal{S}_j^{\alpha }(q_{34}) \mathcal{S}_j^{\nu }(q_{12}) 
\Big)
\bigg]
\nonumber\\
&
\qquad\qquad
+  4C_F \mathcal{S}_j^{\nu }(q_{12}) \bigg[
\frac{q_1^\alpha q_1^\sigma }{(q_1\cdotnew q_{34})^2}
-\frac{2q_1^\alpha q_2^\sigma  }{(q_1\cdotnew q_{34}) (q_2\cdotnew q_{34})}
+\frac{q_2^\alpha q_2^\sigma }{(q_2\cdotnew q_{34})^2}
\bigg]
\Bigg\}
+ (i \leftrightarrow j).
\end{align}
Note that while the expression for $Q_{ijl}$ is not explicitly symmetrized in hard-parton indices, its symmetrization can be straightforwardly achieved.
Unlike the triple soft $q\bar{q}g$ current in similar limits discussed in \cite{DelDuca:2022noh}, the new feature here is the $C_F$ term also contributes to the strongly-ordered soft limit \eqref{Strongly-ordered-limit-qqQQ}.

\section{Conclusions}\label{sec-conclusion}

In this work we have initiated the study of four soft parton emissions. We derived the tree-level current for the radiation of two soft quark-antiquark pairs, $q\bar{q}Q\bar{Q}$, using the eikonal formalism. We also explicitly worked out the square of this current and analyzed the resulting color correlations, which can be categorized into three distinct types: double dipole, dipole, and tripole correlations, similar to those observed in the emission of a soft quark-antiquark pair plus a soft gluon \cite{DelDuca:2022noh}. Our result is a crucial ingredient for investigating infrared singularities at N$^4$LO in perturbative QCD.

\vskip 5pt
Further investigation into soft factorizations relevant to the N${}^4$LO accuracy in QCD computations would be an interesting direction. This involves computing tree-level soft currents for $ggq\bar{q}$ and $gggg$ emissions, one-loop triple and two-loop double soft emissions, as well as splitting amplitudes for the emission of up to five collinear partons, such as \cite{Guan:2024hlf}. While we leave these tasks for future research, we provide some preliminary observations here. First, it is reasonable to expect that the tree-level soft $ggq\bar{q}$ current may share structural similarities with the triple-soft gluon current derived in \cite{Catani:2019nqv}, as a soft quark-antiquark pair can behave analogously to a soft (off-shell) gluon. For the emission of four soft gluons, we can relatively easily construct the tree-level current using in a Berends-Giele-like recursive algorithm. We also find a systematic method to manage color chargers and write the result in an organized form. However, simplifying the squared currents remains challenging. In particular, the algebraic manipulation of color charges in squared currents can be `{\it quite cumbersome}', as described by Catani and collaborators \cite{Catani:1999ss,Catani:2019nqv}. Therefore, developing an efficient algorithm for reducing color charges in squared currents into a set of irreducible correlations would represent a significant advancement.

\section*{Acknowledgements}
We would like to thank Vittorio Del Duca, Claude Duhr, Einan Gardi and Rayan Haindl  for a fruitful collaboration on this topic. ZL is grateful to the Galileo Galilei Institute for the hospitality and the INFN for partial support during the scientific programs on {\it Theory Challenges in the Precision Era of the Large Hadron Collider} (where this work was initiated) and {\it Mathematical Structures in Scattering Amplitudes} (where it was finalized). ZL would also like to express gratitude to Yang Zhang and the Peng Huanwu Center for Fundamental Theory in Hefei for the kind hospitality, during the completion of this work.
This research was supported by the MIAPbP which is funded by the Deutsche Forschungsgemeinschaft (DFG) under Germany’s Excellence Strategy\,--\,EXC-2094\,--\,390783311.
This work was supported partially by the Start-up Research Fund of Southeast University No.\,RF1028624160.
This work was supported partially by the European Union's Horizon 2020 research and innovation program under the Marie Sk\l{}odowska-Curie grant agreement No.\,847523 `{INTERACTIONS}'.

\newpage
\input{ref.bbl}
\end{document}

%% file: ref.bbl
\providecommand{\href}[2]{#2}\begingroup\raggedright\endgroup